\documentclass[conference]{IEEEtran}
\usepackage{cite}

\ifCLASSINFOpdf
  \usepackage[pdftex]{graphicx}
  \usepackage{epstopdf}
  \graphicspath{{./pic/}}
  \DeclareGraphicsExtensions{.pdf,.jpeg,.png}
\else
  \usepackage[dvips]{graphicx}
  \graphicspath{{./pic/}}
  \DeclareGraphicsExtensions{.eps}
\fi

\ifCLASSOPTIONcompsoc
    \usepackage[tight,normalsize,sf,SF]{subfigure}
\else
    \usepackage[tight,footnotesize]{subfigure}
\fi

\usepackage{amssymb}
\usepackage[cmex10]{amsmath}
\usepackage{algorithm}
\usepackage{algorithmicx}
\usepackage{algpseudocode}
\usepackage{amsmath}
\usepackage{array}
\usepackage{mdwmath}
\usepackage{mdwtab}
\usepackage{amsfonts}
\usepackage{url}
\usepackage{color}

\begin{document}
%
\title{HVSTO: Efficient Privacy Preserving Hybrid  Storage in Cloud Data Center}

\author{\IEEEauthorblockN{Mianxiong Dong\IEEEauthorrefmark{1}, He Li\IEEEauthorrefmark{2}, Kaoru Ota\IEEEauthorrefmark{3}, Haojin Zhu\IEEEauthorrefmark{4} }

\IEEEauthorblockA{\IEEEauthorrefmark{1}
University of Aizu, Japan \\
}

\IEEEauthorblockA{\IEEEauthorrefmark{2}Huazhong University of Science and Technology, China\\
}

\IEEEauthorblockA{\IEEEauthorrefmark{3}Muroran Insitute of Technology, Japan\\
}

\IEEEauthorblockA{\IEEEauthorrefmark{4}Shanghai Jiao Tong University, China\\
}
 
}

%


\maketitle

\begin{abstract}
In cloud data center, shared storage with good management is a main structure used for the storage of virtual machines (VM). In this paper, we proposed Hybrid VM storage (HVSTO), a privacy preserving shared storage system designed for the virtual machine storage in large-scale cloud data center. Unlike traditional shared storage, HVSTO adopts a distributed structure to preserve privacy of virtual machines, which are a threat in traditional centralized structure. To improve the performance of I/O latency in this distributed structure, we use a hybrid system to combine solid state disk and distributed storage. From the evaluation of our demonstration system, HVSTO provides a scalable and sufficient throughput for the platform as a service infrastructure.


\end{abstract}


%
\IEEEpeerreviewmaketitle

\section{Introduction}
In cloud data center, virtualization technology brings flexibility and reliability to the cloud service \cite{Winter2009}\cite{Carr2009}.  In Infrastructure as a Service (IaaS) and Platform as a Service (PaaS), since virtual machines (VM) are the main interface to provide cloud service to users \cite{Mell2011}, to protect the VM data is essential to user privacy. In cloud data center, virtualization provides effective data isolations \cite{Vaquero2008}. Storage encapsulation is the main method to isolate VM data in logical level. All data of each VM are encapsulated to one or more disk image files stored in a storage system \cite{Sotomayor2009}.

However, it is not always secure for user privacy by this encapsulation. To support VM management in cloud data center, people use a shared storage in which all data of VM are stored in a uniform storage space. To implement this shared storage, existed works adopt centralized structure that all physical nodes connect to a centralized storage unit like NAS or other storage system  \cite{Vaghani2010}. Even it is convenient for management with sufficient performance by adopting high-end storage devices, centralized structure easily due to a data leakage of all VM in the whole cloud data center when some nodes with the access privilege are comprised. \cite{itani2009}.

To prevent security issues of this centralized structure, access control is used, which allows only the limited part of the cloud data center has the right to access the centralized storage \cite{Wei2009b}. It can prevent some malicious accesses from a compromised node. While the access from physical nodes or the administrator is allowed,  , it is possible that the centralized storage system is compromised to leak VM data in some worst cases.

In this paper, we proposed HVSTO, a distributed storage system for preserving privacy when some storage units are compromised. We design a distributed structure to spread security risks to multiple storage unit. Meanwhile, we design a block mapping for each VM, with which the data of each VM is distributed in each storage unit. A compromised storage unit can only get a part of data on each VM. To preserve privacy, HVSTO splits these data to small blocks and sparsely stored in each storage unit. It is difficult to get information by using parts of data blocks in one or several storage unit.

As the traditional address mapping is not suitable for the distributed storage structure, we design a new tree-like mapping structure in HVSTO. HVSO also splits the mapping data to small blocks and stores this blocks to the distributed storage units sparsely. With this distribution, compromised nodes are hard to get the full metadata of each VM image file, without which it is hard to reorganize the blocks to data. Meanwhile, with this mapping structure, a VM image is organized by a version tree with extraordinary performance snapshot.

With the distributed structure which provides enough concurrency for virtual machines, the latency is sometimes difficult to prevent since the block mapping and network transferring. In HVSTO, with a small block distribution, it is hard to ignore the I/O latency from the distributed structure. Traditional method to decrease this latency is used a high-end network. In HVSTO, we adopt a low-cost method that adopts a hybrid structure to combine the local solid state disk (SSD) and the distributed structure. Although the maximum bandwidth of the SSD device is limited, the high IOPS performance is enough for supporting multiple VM concurrently access. In HVSTO, local SSD stores the metadata, shared image data and parts of branch data of each VM above it. From the evaluation, the concurrency performance is improved by this hybrid design.

The main contributions of this paper are summarized as follows.
\begin{itemize}
\item First, we proposed HVSTO, a shared storage system with a distributed structure to preserve privacy even parts of storage units are compromised. 
\item Second, based on this distributed structure, we design a new mapping structure for better privacy preserving and high efficiency. 
\item Last, with the small block distribution, we design a hybrid structure that combines the local SSD and distributed storage units to provide better storage performance.
\end{itemize}


The rest of this paper is summarized as follows. We discuss the threat model and design themes in Section \ref{sec:design}. The details of design and implementation are discussed in Section \ref{sec:implementation}. In Section \ref{sec:eva}, our evaluation is divided in two parts: first we evaluate the overload on a single node and the concurrency performance of a small cluster. The last sections offer some concluding thoughts and future works.

\section{Problem Statement}
\label{sec:design}

\subsection{The System and Threat Model}
\begin{figure}[!h]
\centering
\includegraphics[width=0.95\linewidth]{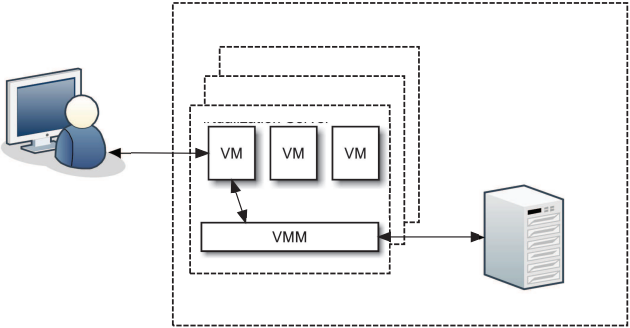}
\caption{The threat model of the VM storage in cloud data center.}
\label{fig:threat_model}
\end{figure}

As shown in Fig. \ref{fig:threat_model}, we consider storage in a cloud data center network involving four different entities: the user, who considers their data are stored in VM; the virtual machine (VM), who sends or receives storage request to the virtual storage device; the virtual machine monitor (VMM), who manages the virtual storage devices and transferring storage I/O request to the center storage system as I/O request to the image files; center storage system, in which all VM data are stored.

In general case, each virtual machine has a virtual storage device for I/O request in cloud data center. The virtual storage device is encapsulated to a virtual disk image that is stored in the storage system. When the user does something to their operation system due to some I/O requests to the filesystem in VM, VM sends the I/O requests to the virtual storage device. VMM receives these requests from VM through the virtual storage device and forward the request to the network storage interface which connects the storage system. The storage system processes these requests to the physical storage devices. 


In this model, a potential threat is all VM data are stored in a centralized storage system logically in which all data are accessible. If some compromised nodes in the cloud data center get the access permission to this storage system, it is not difficult to get all VM image files. Considering the limitation of the hardware performance, VM image files are hard to store in the storage system with encryption. As a result, this malicious access can easily analyze the detail of VM image file, which leads to a leakage of user privacy.


\subsection{Design Goals}
We define three different design themes as following to reduce the leakage risk of user privacy without decreasing the performance
\subsubsection{Disributed Shared Storage}
As mentioned before, centralized storage is the main defect to the threat. Even using access control or other defense methods, it has an obligation to open the access permission to some nodes like virtualization servers to execute VM and the manage nodes who control the behavior of VM like migration, snapshot or other essential operations. It is hard to negative the probable threat that a node with the access privilege to the storage system is compromised. A feasible method to vanish this threat is using a different storage structure. As a result, we use a distributed structure to store VM images in HVSTO.

Although dividing VM data to multiple storage units reduces the leaked data in the worst case, it harm to the scalability of resource management. It is hard to schedule storage resources except moving whole VM image data from one storage unit to another. In HVSTO, we design a shared space to provide a same storage space to each virtualization server while VM data are sliced to small blocks and distributed sparsely in multiple storage nodes. Therefore, it becomes almost impossible to get the VM data though a pile of small discrete blocks on one or several compromised storage nodes.


\subsubsection{Efficient Mapping}
Conventional virtual machines storage solutions like VMDK\cite{VMDK2007} or QCOW\cite{QCOW2007} were using a "chain" mapping to realize benefits of virtual disks. In this "chain", each version of the original disk file will store the increment and a point to the former version means an I/O request to source disk data on the $N$ version will jump $N$ times. It is not an obvious problem with existed high-end storage system which provides adequate performance includes low latency access. In HVSTO, this version control is not suited to the distributed storage that the access between storage node and virtualization server is slower than high-end storage system.

Meanwhile, actually, the "chain" mapping is based on the sequential storage which is easy to locate the rest of the data with one address of the block in an image. If the compromised node gets one or more block addresses, the attacker can get more VM data easily. In HVSTO, instead the "chain" mapping, we design an efficient direct index metadata for mapping a virtual block to a physical block. Based on these structures, with the generally disk image reconsidered on a map of non-sequential blocks, using parts of mapped addresses is hard to get other VM data.

\subsubsection{Hybrid structure}
We design a hybrid storage structure combined by a local solid-state disk equipped with every node of virtualization cluster and the distributed storage. As mentioned before, since serial access module of the general mechanical disks, their limited concurrency performance could not support enough virtual machines. Storage service devices increase could remit the pressure of concurrency access from virtual machines in PaaS. Another efficient method is using the local storage devices equipped in virtualization nodes.

In most virtualization implementations, it is necessary to equip enough capacity storage device for installation some required components like a virtual machine monitor(VMM)\cite{Barham2003}. We combine these devices with the shared distributed storage for increasing the data I/O throughput. Consider throughput from tens of virtual machines in a single virtualization node, we choose the solid state disk that a flash based storage device without any mechanistic structure. With the characteristics of SSD, the random I/O throughput is thousands of times than the traditional hard disk. This feature meets the demand of storage virtualization\cite{Agrawal2008}.

\subsection{Security Analysis of Distribution}
To better understand the better security of the distribution structure, we analyze the possibility of user data leakage in a small case. 

Firstly, we introduce a small case in cloud data center for our analysis. There a small cloud data center with $M$ VMs and $H$ private data existed in these VMs. To a private data $d_i$, we use $l_i$ to denote the data size. 

Then we define the $p_i$ to denote the possibility of the leakage of data $i$. Therefore, we get the possibility $p_i$ when the storage system is compromised in a centralized storage system. Oblivious, this possibility is $100\%$ since the $l_i$ size of data are obtained by scan the whole storage system.

After that, we adopt our distributed design of storage system. In this storage system, all image files are sliced to small blocks and distributed to $N$ storage nodes. We define $s$ to denote the size of each block. Considering a simple random distribution, we get the possibility of a block stored in one of $n$ specific storage node is $\frac{n}{N}$. 

Therefore, to get a private data $i$, the malicious application needs to get $n_i$ blocks to recovery the content of these data. The definition of $n_i$ is shown in (\ref{eq:n_i}).

\begin{equation}
\label{eq:n_i}
n_i = \begin{cases}
 \frac{l_i}{s} & \text{ if }  l_i \geq s  \\ 
 1 & \text{ if } l_i < s  
\end{cases}
\end{equation}  

Considering the distribution of each block is a separate event, as shown as in (\ref{eq:p_i_1}), we easily get the possibility $p_i$ that the private data $i$ are leaked when $n$ storage nodes are compromised. 

\begin{equation}
\label{eq:p_i_1}
p_i = \begin{cases}
 (\frac{n}{N})^{\frac{l_i}{s}} & \text{ if }  l_i \geq s  \\ 
 \frac{n}{N} & \text{ if } l_i < s 
 \end{cases} 
\end{equation}

Therefore, we get the total size $P$ of leakage data when $n$ storage nodes are compromised as shown as in \ref{eq:total}.

\begin{equation}
\label{eq:total}
P = \sum_{i=1}^{H}p_i l_i = \sum_{i=1}^{H}(\frac{n}{N})^{n_i}l_i
\end{equation}

To evaluate the security of our system, we take a simulation based on several traces of real world user data. We describe the detail of this simulation in Section \ref{sec:eva}.

\section{Design and Implementation}
\label{sec:implementation}

\subsection{System Architecture}

\begin{figure}[!ht]
\centering
\includegraphics[width=2.5in]{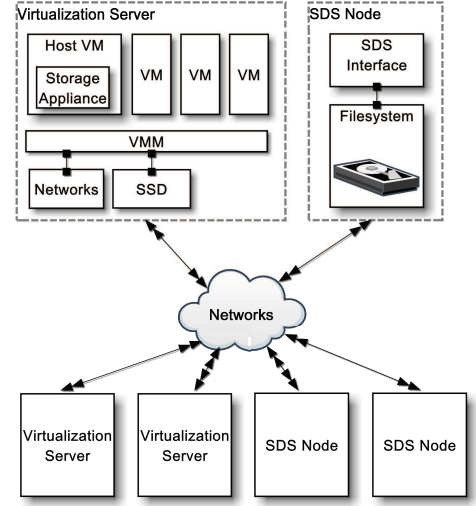}
\caption{Hivo consist of the storage appliance in virtualization servers and the shared distributed storage(SDS)}
\label{figure:HVSTO}
\end{figure}
As shown in Fig. \ref{figure:HVSTO}., HVSTO consist of shared distributed storage (SDS) and local storage appliance. SDS is a group of commercial computers equipped generally storage devices and the virtual block interface daemon. Local storage appliance is a toolkit for redirecting the I/O requests from virtual machines to the SDS and management of local SSD cache.

HVSTO provides a virtual disk interface for each VM executing upon the virtualization servers accessing the storage transparently. HVSTO uses a block index structure instead of the traditional file abstraction to manage each virtual disk image. As the general file storage, to access data in HVSTO, it is necessary to inquiry the metadata stored in the SDS. Metadata and data are both managed by the local storage appliance.

The local appliance on the host VM gets all I/O requests from virtual machines then transfer these requests through general TCP/IP networks to the storage nodes in SDS or access the local SSD if the destination data of these requests were cached. Since the isolation of virtualization, these procedure is agnostic to the operation system in each VM.

The virtual block interface daemon accepts the network package from the local storage appliances and processes these requests to final storage devices. For the design of a distributed structure, in virtual block interface daemon, we design three steps of processing I/O requests from virtualization servers. First, the interface daemon calculate the destination nodes of the request data block. Second, the interface daemon transfer each request to the destination nodes. Last, daemon process requests to the blocks contained in the local node.

\subsection{Meta-data}
\begin{figure}[!ht]
\centering
\includegraphics[width=3.5in]{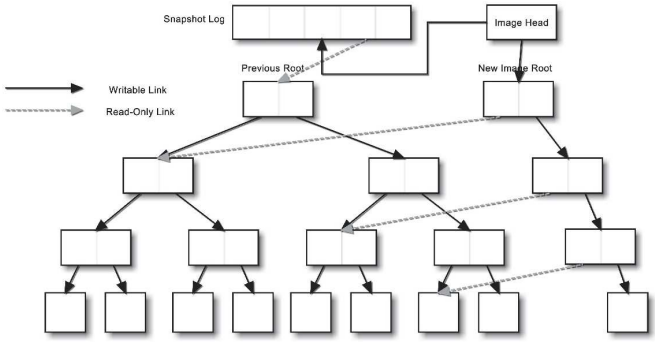}
\caption{The Meta-data Structure and Snapshot Procedure of HVSTO}
\label{figure:metadata}
\end{figure}

In some newer design, a block direct index of virtual disks was used for a better performance in virtual machines storages. By this method, each block in each version of virtual disks is mapped to the physical block in an index data structure. It is avoided that the linear increased delay of processing I/O request in a deep version tree. Since the benefit on performance of the block direct index, we choose a similar way in HVSTO. In a procedure of sending each request of a virtual block to the mapped physical one, we found the time overhead occur in the seeking disk for index data both direct index or \"chain\" index and repeatedly reading disk when using the letter is the reason of linear performance decrease. But after disk seeking, the data read time is unremarkable. So we adopted a B+tree like structure from database systems to index the virtual block to physical.

The nodes in tree are in various sizes as mentioned in Fig. \ref{figure:metadata}. We take a three depth B+tree for mapping virtual block to the space in distributed storage.
Based on this B+tree, for the disk image snapshot, we implement a copy-on-write mechanism. When a snapshot is taken, a new image root is created and a read-only link will be set to leaf of the previous root. When virtual machine generates a new block, HVSTO will construct a writeable link and the non-leaf nodes to this block. The other link of new non-leaf nodes will be set read-only to the corresponding child in the previous B+tree structure. And the newest image root will be set writeable and the previous tree will be set read-only in the snapshot log. With this efficient snapshot, in HVSTO, we design a rewrite avoidance in I/O processing. When the VM wants to change the data in its image, HVSTO takes a snapshot of the current image and generate a new version to store the new data. The rewrite avoidance significantly simplifies the processing of the rewrite request. We also design and implement a garbage collection to remove the automatically generated version when low load periods.

\subsection{Data in SSD}
HVSTO has three different types of blocks, metadata, system image and activity data. Since the limited space of local storage, SSD stored a part of data of virtual machines on the virtualization machine. We divide the local SSD storage space to three distinguished parts for storing this different data and implement different replacement strategies. In the SSD device, we set 25\% storage space for caching the metadata, 50\% for system image and the rest for active data.

With rewrite avoidance, in the period of system service, the meta-data size will increase since more and more snapshot of VM file systems. We modify the Least Recently Used (LRU) cache replacement strategy to the metadata in two stages. First, scan the latest version of disk images used by all upon virtual machines and label these blocks cache index as protected. Second, remove the last element unprotected of the LRU queue. Since we set each image index should be accessed by a single VM strictly, the cached metadata update only takes place in write-back.

To release the concurrent access pressure during some determinable periods like booting VM, local SSD caches parts of source VM images. The contents in these images are only changed by the administrator in system upgrading or exceptional cases. However, with various VM existed in the whole cloud, the storage space in SSD is limited to store all source image. When some VMs migrate in or boot up, it is needed to updata this part of SSD space. We adopt LIRS cache \cite{Jiang2002} to manage this space to guarantee as many essential data stored in local as possible during the peak load.

The active data space caches the read/write data of active VM.  These VMs have their read/write cache space. For the read cache, we implement prefetch and replacement strategy refereed by the Linux kernel file cache with a larger prefetch window. HVSTO provides a 100MB write cache in SSD for each active VM. As mentioned before， to implement rewrite avoidance, HVSTO update the metadata more frequently. To reduce this update, new version is created until the write cache full or VM is saved/migrated.

\section{Evaluation}
\label{sec:eva}
As we described in the past sections, HVSTO has some special features supports VM storage in cloud environment. To confirm these features, we evaluate HVSTO in two different perspective. 
\subsection{Data Privacy Simulation}
\begin{figure}[!ht]
\centering
\includegraphics[width=3.5in]{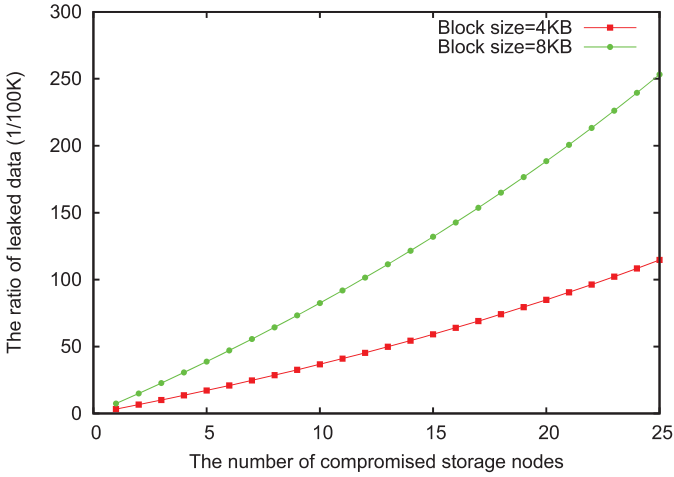}
\caption{The user data leakage ratio when 1 to 25 are compromised.}
\label{figure:security}
\end{figure}

In this section, we firstly evaluate the security efficiency of HVSTO by simulation with some trace of real word user data. Then we take some tests to evaluate the performance of HVSTO. We trace user files in \\My Documents of Windows 7 of 10 student computers in our laboratory and record the file name and size (About 20.86GB totally). We consider there 100 storage nodes of the small cloud data center and the block size is 4KB and 8KB. Then we put these records to calculate the leaked data ratio these 10 users if they are using cloud services instead their personal computers. We calculate the leakage ratio from 1 to 25 nodes are compromised. 

As shown in Fig. \ref{figure:security}, we find that the leakage ratio of user data after storage system is intruded is very small with the design of HVSTO. When one storage node is compromised in HVSTO with 4KB block size, the leakage ratio is only 0.0033\% or it is possible to leak about 65.6KB user data. While the block size becomes 8KB, the possible leakage ratio is 0.0074\%. If more nodes are compromised, the possilbe leakage ratio becomes higher. When $\frac{1}{4}$ nodes are compromised, it is possible to leak about 
2.45MB user data, which means leakage is more than 0.11\% with 4KB block size while the possible leakage ratio becomes 0.25\% with 8KB block size. Even though there are a few of data could be leaked when some storage nodes are compromised, HVSTO is much better than the single storage system which has 100\% leakage ratio after it is compromised.

\subsection{Testing Configuration}
To measure the efficiency of HVSTO, we test HVSTO in two different persecutive. We design a micro benchmark to find the effects brought by each design and implementation.

All test is taken in a 7 nodes cluster with commercial blade servers. In this cluster,  each node equips two 1.6Ghz Intel Xeon E5310 quad core processors, 4 GByte of  RAM, an Intel 40GB X25-V SSD, a 500GB HDD and an Intel e1000 GbE network interface cards. All servers are connected by a TP-LINK TL-SF1016D switch. As the comparison system, we choose one node with NFS protocol \cite{Callaghan1995} to provide the centralized storage service. To the micro benchmark, we choose two nodes that one node as the virtualization server and the other as the storage node to test the I/O latency. To the overall performance test, all 7 nodes are used. In all test, the software environment is the same. We install XCP 1.6 with CentOS 6.3 on each node and use Fedora 15 without GUI as the OS of guest VM.

\subsection{Micro Benchmark}
In this persecutive, based on this mapping structure and hybrid design, we test the I/O latency brought by distributed design with mapping structure and hybrid design. Then, we measure the performance degradation of the distributed structure than general local storage devices. 
%
%
%

\subsubsection{I/O Latency}
\begin{figure}[!ht]
\centering
\includegraphics[width=3.5in]{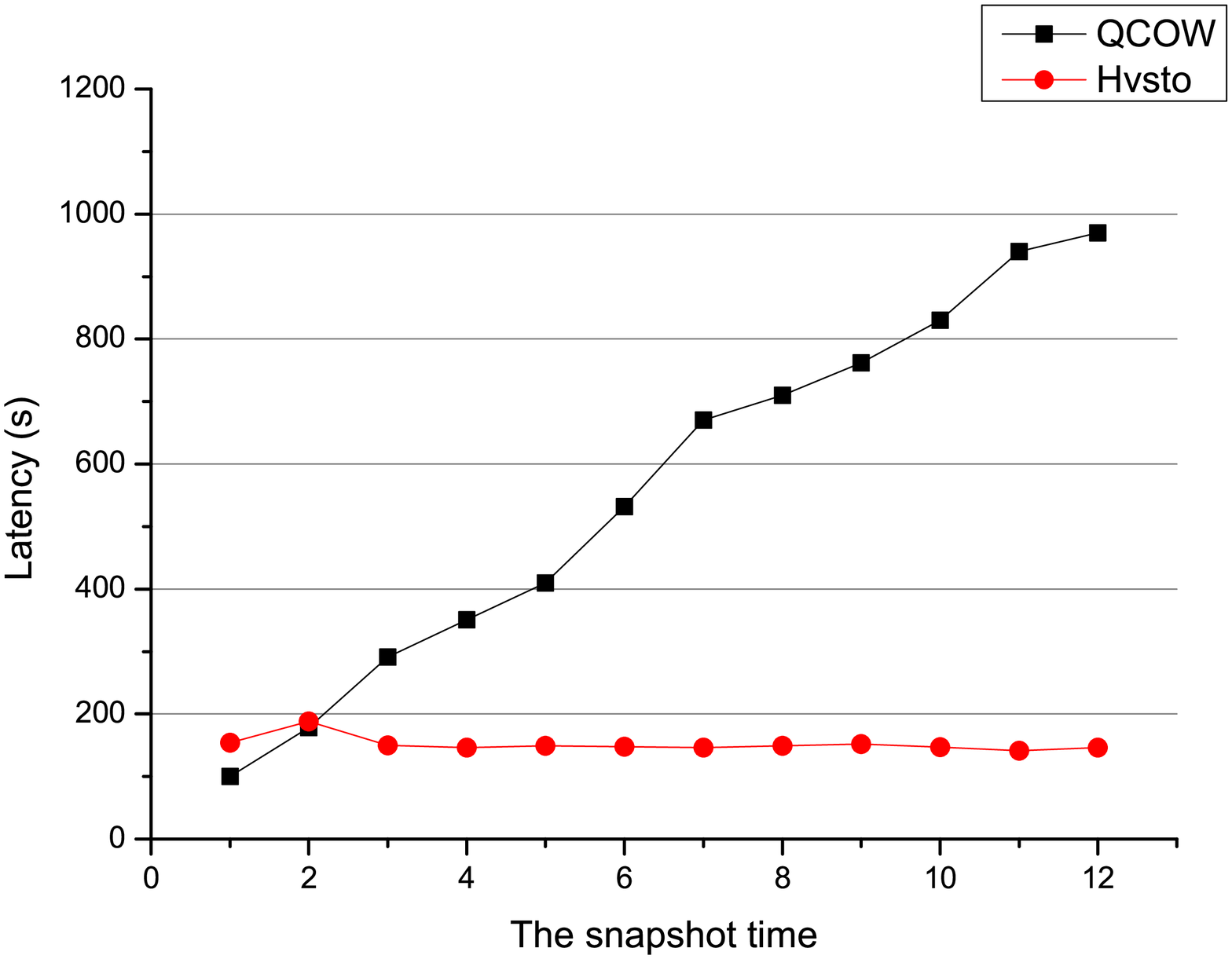}
\caption{The read latency after snapshot of HVSTO and general QCOW format image}
\label{figure:snapshot}
\end{figure}

As mentioned in Section \ref{sec:design}, with the distributed structure, the I/O latency in HVSTO is bigger than general storage system. We adopt a Hybrid structure to optimize this latency. Meanwhile, in our design themes, we describe the mapping structure in HVSTO provides better snapshot performance to support some advance features for the virtualized storage. 

Therefore, we measure the I/O latency of HVSTO with version control. We describe the test steps as following. Initially, we choose a VM image and take a snapshot to this VM image in HVSTO. We set this snapshot as a version of VM image. Then, we run a VM with this version and test the latency by recording the average I/O latency of reading same 2MB data in 5 times. After that, we take snapshot to this version to generate a new version and repeat the same test. In latency test, we repeat the test steps 10 times and get the test result from 10 versions of VM image. As a comparison, we take the same test on NFS. We put the same image in NFS and take snapshot on this image by QCOW.

From the result in Fig. \ref{figure:snapshot}, in the source image, with the overload of HVSTO, the latency is bigger than NFS. In the image after 2 times snapshot, the latency of HVSTO is near the QCOW. With the cached data in local SSD, we find the latency is decreased in the image after 3 times snapshot. With more snapshot, the latency of QCOW snapshot is linear increased and HVSTO is almost stable value. From the result of latency test comparing with the NFS, we consider the I/O latency increased by HVSTO is not obvious. In addition, with the mapping structure and hybrid design, to the original QCOW, HVSTO performs better I/O latency of the image after more 3 times snapshot to the source image.

\subsubsection{Throughput}
\begin{figure}[!ht]
\centering
\includegraphics[width=3.5in]{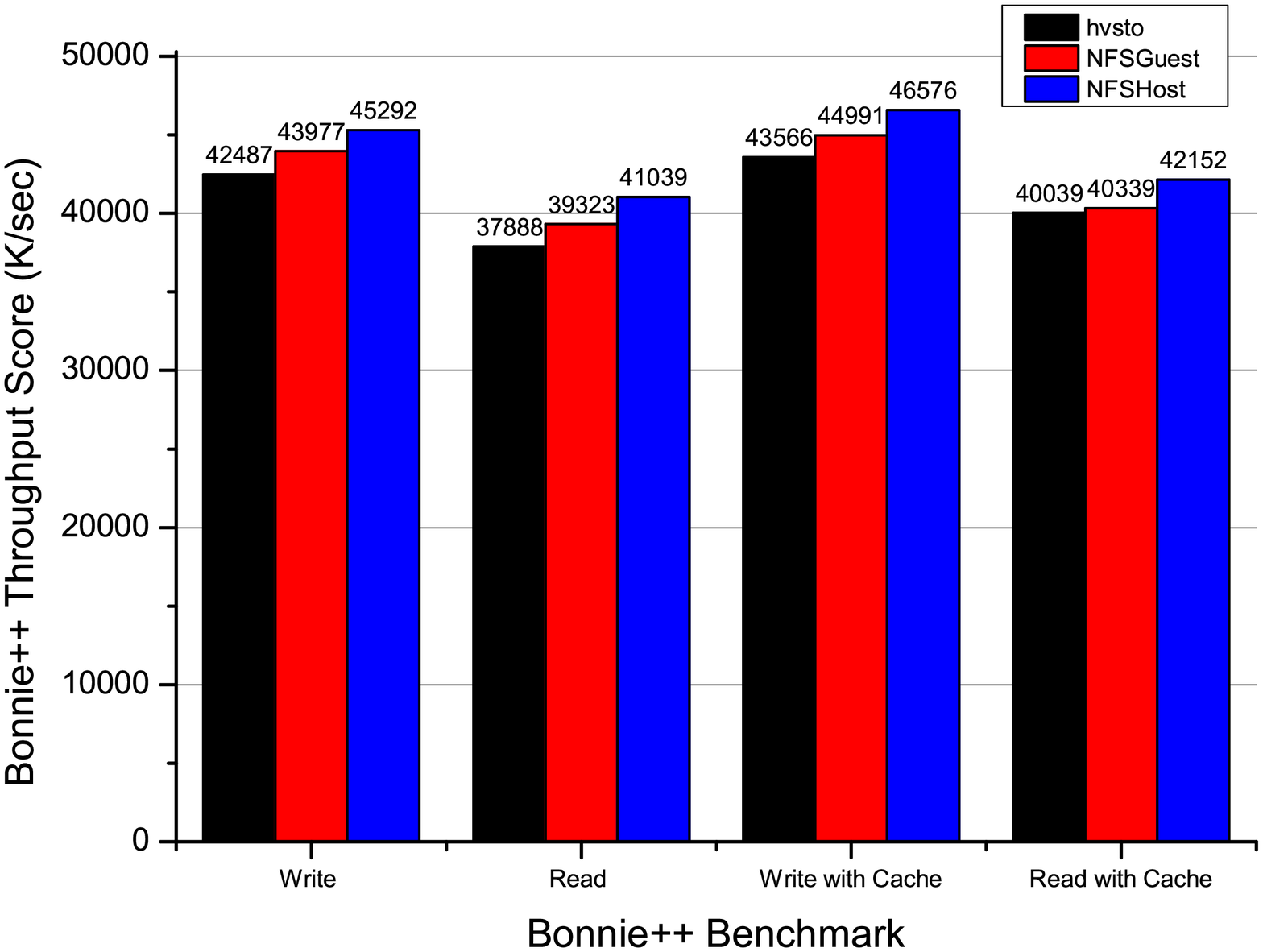}
\caption{The Bonnie++ Output results on different storage devices}
\label{figure:bonnie}
\end{figure}

Since our implementation of virtual block mapping, there is a performance degradation from the original network filesystem. We test the throughput to find out the overload by the HVSTO design. We choose bonnie++\cite{coker2001bonnie++} as the mainly benchmark application to measure the general sequential I/O performance. We run a VM in HVSTO and excute bonnie++ in this VM. As a comparison to find the performance degradation by virtualization and HVSTO, we run the test in a physical node with native NFS and a VM with NFS. As shown in Fig. \ref{figure:bonnie}, the NFSGuest means VM image is stored in NFS and the NFSHost means native NFS. 

From the test result, compared with the performance of native NFS, we find HVSTO has a degradation at 6\% for write throughput performance and 7\% for read throughput while the VM with NFS has a degradation of 3\% for write and 4\% for read. With the filesystem cache, HVSTO has a better result that the degradation is 6\% for write throughput and 5\% for read. To the VM with NFS, the degradation is 3\% and 4\%. Since the result with cache is better to measure the performance of storage system in practical usage, we consider that the overload to a VM in HVSTO design is not obvious with the original  virtualized disk image.

\subsection{Performance Test}
In this part, we test the concurrency access performance which is essential to a large scaled cloud data center. Then, to indicate the efficiency of the distributed structure, we find the system scalability by test the system throughput with the different number of storage nodes. 

\subsubsection{Concurrency Access}
\begin{figure}[!h]
\centering
\includegraphics[width=3.5in]{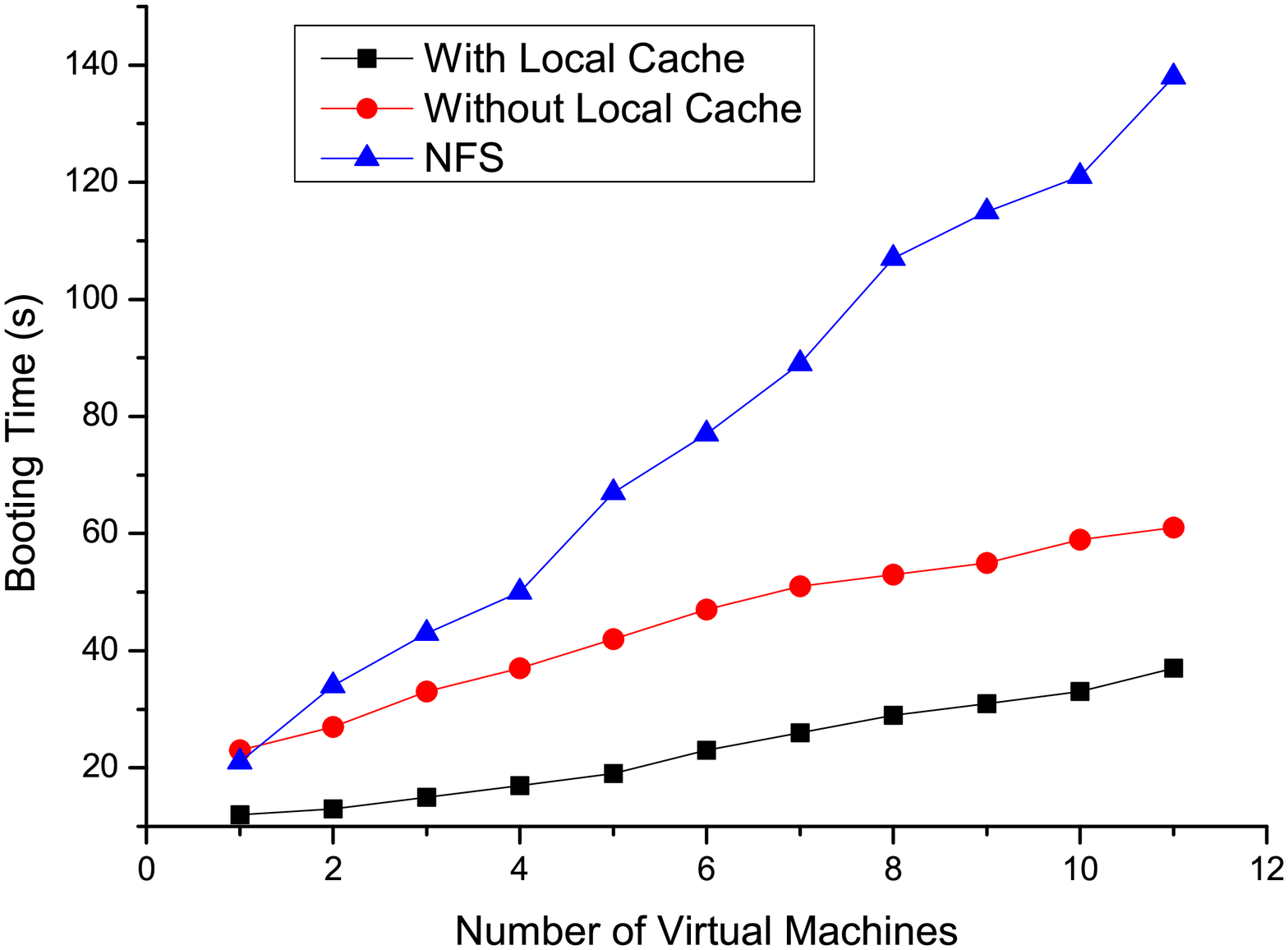}
\caption{The Bonnie++ Output results on different storage devices}
\caption{The booting time with system image cache is better than NFS and HVSTO without cache}
\label{figure:booting}
\end{figure}
We test the concurrency performance through reappearing the peak load happens. Boot storm \cite{Soundararajan2010} is a common peak load in existed cloud data center. We choose multiple VM booting in the same time as the test scenario. We boot 1 to 11 virtual machines on one virtualization server in HVSTO with or without local cache. We also measure the same test using NFS as a comparison. Then we record the time from the beginning to the end. We use a shell script to boot these VM simultaneously and sends a signal after all VM booted. We record the system time when the first VM begins booting and the last VM finishes booting. And the interval of this two time is the booting time.

From the results shown in Fig. \ref{figure:booting}, we find that HVSTO perform much shorter booting time than NFS. More additional time is cost with NFS when booting more VM. Additionally, with the local cache, the time is shorter than HVSTO without local cache. Therefore, we consider that multiple nodes in HVSTO provides higher concurrent performance than single NFS node and the performance is increased obviously by hybrid structure.

\subsubsection{Scalability}

\begin{figure}[!h]
\centering
\includegraphics[width=3.5in]{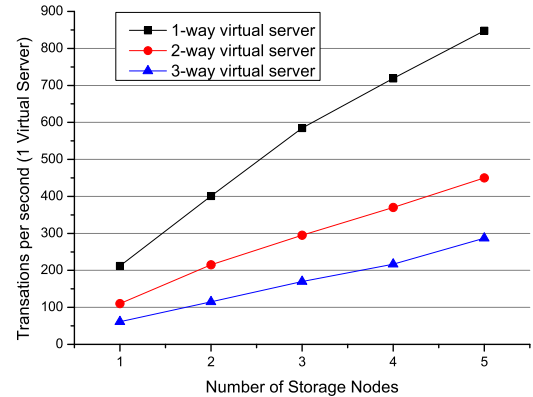}
\caption{The Postmark results on different storage devices}
\label{figure:postmark}
\end{figure}
We obtained eight blade servers whose the equipments mentioned before to test the scalability of HVSTO in clusters. In these eight blade servers, we select 3 nodes as the virtualization servers and other 4 as the storage nodes. We execute postmark \cite{Katcher1997} in each virtual machine concurrently and there are 4 VM on every virtualization server. We record the total performance of all VM in one virtualization server (virtual server) during serving 1-way, 2-way and 3-way virtualization servers. 

From the result in Fig. \ref{figure:postmark}, the plot of the postmark transaction performance reflects that the throughput capacity of the storage network is increased with the storage nodes. When serving 1-way virtualization server, the transactions per second are from 63 to 287 with increasing the number of storage nodes from 1 to 5. When serving 2-way servers, the result is from 109 to 456 and the result of 3-way servers is 205 to 842. From previous fileystem benchmark, the postmark performance of ext3 filesystem on general hard disk is approximately 300 transactions per second, we consider that HVSTO can provide same performance when the number of virtualization servers to storage nodes is more than 3:5.

\section{Conclusion}
To solve the privacy threat brought by centralized storage structure in the cloud data center, we propose HVSTO, a distributed storage system for virtual machines. With a specified design of mapping structure, HVSTO provides better privacy protection and efficient snapshot than original VM image structure. To solve the performance degradation of distributed structure, we adopt a hybrid structure that keep more VM data in local SSD storage to reduce the network interactions. We implement three types of cache in this local SSD storage and the evaluation indicate this cache increase the performance of HVSTO. Considering HVSTO is just a demonstration implement, we will continue to improve the design of HVSTO include better block distribution algorithm, strict access control to virtualization server and scheduling storage resource dynamically.


\section*{Acknowledgment}
This work is partially supported by JSPS KAKENHI Grant Number
25880002, JSPS A3 Foresight Program.




\bibliographystyle{IEEEtran}

%

\end{document}